\documentclass[pre,superscriptaddress,showkeys,showpacs,twocolumn]{revtex4-1}
\usepackage{graphicx}
\usepackage{latexsym}
\usepackage{amsmath}
\begin {document}
\title {Phase transitions in Ising models on directed networks}
\author{Adam Lipowski}
\affiliation{Faculty of Physics, Adam Mickiewicz University, Pozna\'{n}, Poland}
\author{Ant\'onio  Luis Ferreira}
\affiliation{Departamento de F\'{i}sica, I3N, Universidade de Aveiro,  Portugal}
\author{Dorota Lipowska}
\affiliation{Faculty of Modern Languages and Literature, Adam Mickiewicz University, Pozna\'{n}, Poland}
\author{Krzysztof Gontarek}
\affiliation{Faculty of Physics, Adam Mickiewicz University, Pozna\'{n}, Poland}
\begin {abstract} We examine Ising models with heat-bath dynamics on directed networks. 
Our simulations show that Ising models on directed triangular and simple cubic lattices undergo a phase transition that most likely belongs to the Ising universality class. On the directed square lattice the model remains paramagnetic at any positive temperature as already reported in some previous studies. We also examine random directed graphs and show that contrary to undirected ones, percolation of directed bonds does not guarantee ferromagnetic ordering. 
Only above a certain threshold a random directed graph can support finite-temperature ferromagnetic ordering. Such behaviour is found also for out-homogeneous random graphs, but in this case the analysis of magnetic and percolative properties can be done exactly. Directed random graphs also differ from undirected ones with respect to zero-temperature freezing. Only at low connectivity they remain trapped in a disordered  configuration. Above a certain threshold, however,  the zero-temperature dynamics quickly drives the model toward a broken symmetry (magnetized) state. Only above this threshold, which is almost twice as large as the percolation threshold, we expect the Ising model to have a positive critical temperature. 
With a very good accuracy, the behaviour on directed random graphs is reproduced within a certain approximate scheme.
\end{abstract}

\maketitle

\section{Introduction}
The Ising model is certainly one of the most intensively studied models in statistical mechanics. It provides a wealth of possible behaviours and serves also as a testing ground of various approximate or numerical methods. In addition to describing the properties of magnets or a liquid-gas phase transition, it also finds more exotic applications in socio- or econophysics~\cite{stauffer}. This is mainly due to the fact that our decisions (buy or sell), preferences (republicans or democrats) or attitudes (optimistic, pessimistic) very often can be (approximately) represented in terms of some two-state variables. Thus,  certain aspects of a social system or a financial market can be modeled using some kind of Ising-like systems. Not surprisingly, many versions of this model were proposed and studied. An interesting class of models consists of Ising models on directed networks. Since social or economical links are often unsymmetric, an interest in directed models stems mainly from such nonphysical applications.

The Ising models on directed lattices are much different from their undirected versions even at the very basic level. Indeed, for the latter ones a detailed balance is satisifed and in the steady state they are described  by the well-known equilibrium Hamiltonians. For directed Ising models the detailed balance is not satisifed and such models have only dynamical meaning~\cite{sanchez} (nevertheless, in some cases one can find that the steady state is described by certain equilibrium Boltzmann-Gibbs distributions~\cite{godreche}).

Since Ising model plays a fundamental role in statistical mechanics, it would be certainly desirable to understand the behaviour of its  versions on directed networks.
Monte Carlo simulations of Lima and Stauffer suggest that on directed regular lattices the behaviour is much different from its undirected counterpart~\cite{lima2006}. In particular, the square (directed) lattice Ising model seems to be disordered at any positive temperature and such behaviour is suggested also for higher dimensional models. 
Their conclusion in the case of the square lattice is confirmed with the (exact!) analysis by Godr\`eche and Bray, who confirmed the absence of finite temperature ferromagnetism~\cite{godreche}.
However, their analysis of the Ising model on the directed Cayley tree suggests that three- and higher-dimensional models might have a sufficiently large branching number to support a finite-temperature ferromagnetic ordering. 

Taking into account interdisciplinary applications of the Ising model, it is interesting to examine its behaviour also on heterogenenous networks. Such networks include random graphs, scale-free networks or small worlds and one expects that they provide more realistic description of social or economic systems~\cite{albert}. While undirected Ising models on such networks are relatively well understood~\cite{mendesreview}, the behaviour of their directed counterparts remains to a large extent unknown. Some studies predict that the Ising model on a directed random graph undergoes a phase transition, which should be basically similar to its undirected analogoue~\cite{limasumour}. However, in the case of an Ising model on scale-free networks considerable differences between directed and undirected versions were reported. In particular, while for the Ising model on undirected Barab\'asi-Albert network
the critical temperature increases logarithmically with the network size~\cite{aleksiejuk,dorog}, the directed version most likely remains disordered at any positive temperature~\cite{sumourshabat}. Let us notice, however, that Barab\'asi-Albert network is only a particular case of scale-free networks. Ising model on such systems exhibits a rich critical behaviour, but so far mainly the undirected version was examined \cite{mendesreview}.

Trying to understand the behaviour of Ising models on directed networks, we might perhaps find it instructive to compare it with some other phenomena, which take place on such networks. The simplest and most natural candidate is percolation. In the case of undirected random graphs, it was shown that the emergence of a spanning cluster  coincides with the emergence of ferromagnetic solutions of the corresponding Ising model~\cite{replica,dorog}. In other words, percolating cluster provides a sufficiently strong support for the finite-temperature ferromagnetic ordering. It would be interesting to examine whether models on directed networks have a similar property.

In the present paper we examine the Ising model on several directed lattices. Our simulations on regular lattices show that Ising models on triangular and simple cubic lattices have a finite critical temperature and the critical point most likely belongs to the (equilibrium) Ising model universality class. This conclusion is based on the values of two critical exponents. Moreover, simulations confirm that on the directed square lattice the Ising model remains disordered at any positive temperature.
We also analyse the Ising model on directed heterogeneous random graphs and show that in this case the emergence of a percolating cluster does not guarantee a ferromagnetic ordering. Such behaviour is different from the behaviour of undirected random graphs. Additionally, we analyse the zero temperature freezing and show that also in this case some differences between directed and undirected graphs are present.
Moreover, we show that the behaviour of the Ising model on heterogeneous random graphs to a remarkable accuracy can be reproduced within an approximate scheme that assumes some simplified form of a certain probability distribution of spin variables. Finally, we discuss the behaviour of the Ising model on a class of directed random graphs subject to the constraint that the number of out-links is constant. In this case both magnetic and percolative properties can be analysed exactly.

\section{Ising model on regular directed lattices}
In the present paper we examine the Ising model that is a collection of spin variables $s_i=\pm 1$ located at the sites $i=1,2,\ldots, N$ of a certain directed lattice. 
The spin variables evolve according to the heat-bath dynamics; at each elementary step, we select randomly a site $i$ and set the corresponding variable $s_i=1$ with probability
\begin{equation}
r_i=\frac{1}{1+\exp(-2h_i/T)}, \ \ \ \  h_i=\sum_{k_i} s_{k_i},
\label{heat-bath}
\end{equation}
while with probability $1-r_i$, we set $s_i=-1$.
The parameter $T$ might be interpreted as the temperature and the summation in Eq.~(\ref{heat-bath}) includes all sites $k_i$ for which there is an out-link from $i$ to $k$. A unit of simulation time ($t=1$) is defined as $N$ elementary steps (i.e., during a unit of time each spin is updated once, on average). Starting from a certain initial configuration, we let the system to reach the stationary state and then measure the magnetization $m=\frac{1}{N}\langle\sum_i s_i\rangle$ and its variance $\chi=\frac{1}{N}\langle (\sum_i s_i -\langle \sum_i s_i \rangle)^2 \rangle$, where $\langle ... \rangle$ denotes the time average during Monte Carlo simulations. Using the relation which is valid for equilibrium systems, we will refer to the variance $\chi$ as a the susceptibility, even though, as we mentioned, our system is non-equilibrium. However, we will present some evidence that $\chi$ indeed behaves similarly to the  susceptibility in equilibrium models.

\subsection{d=2 square lattice}
\begin{figure}
\includegraphics[width=\columnwidth]{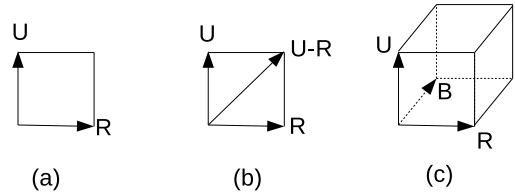}
\caption{Orientation of bonds on square (a), triangular (b) and simple cubic (c) lattices. The triangular lattice is topologically equivalent to a square lattice with one diagonal.}
\label{lattice}
\end{figure}

\begin{figure}
\includegraphics[width=\columnwidth]{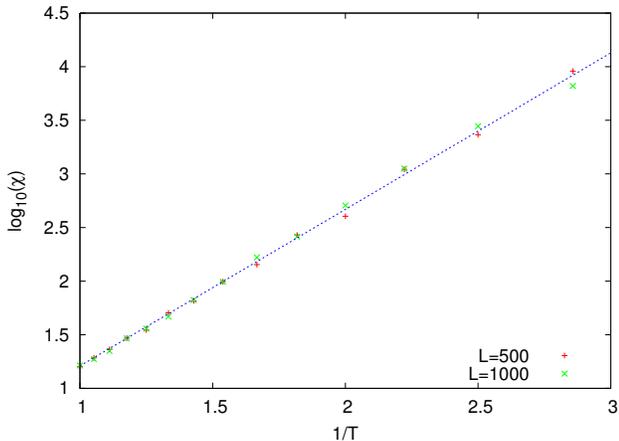}
\caption{(Color online) Logarithm of the susceptibility $\chi$ as a function of $1/T$ for the directed square lattice model. A nearly linear dependence supports the relation $\chi \sim \exp(a/T)$, where $a>0$ is a certain constant.}
\label{suscsq}
\end{figure}

As a first example we analysed the Ising model on a directed square lattice of linear size $L$ ($N=L^2$). At each site links point toward Up and Right neighbours, i.e., these are the only neighbours that have an influence on a given site (Fig.~\ref{lattice}a). The same orientation of bonds was used in some previous works on basically the same model~\cite{lima2006,godreche}. And similarly to the previous works, our simulations suggest that the model does not show ferromagnetic ordering most likely at any positive temperature. We made simulations for system sizes up to $L=10^3$ and even for very low temperatures we did not observe a stable magnetization (instead, the system frequently jumps between the states of positive and negative magnetization). The temperature dependence of the susceptibility $\chi$ provides an additonal confirmation of the lack of ferromagnetic ordering (Fig.~\ref{suscsq}). Indeed, the Arrhenius behaviour   $\log(\chi)\sim 1/T$, over nearly three decades of~$\chi$, suggests the absence of a finite-temperature ferromagnetic phase (a similar behaviour is obtained, e.g., for the $d=1$ equilibrium Ising model). 

\subsection{d=2 triangular lattice}
\begin{figure}
\includegraphics[width=\columnwidth]{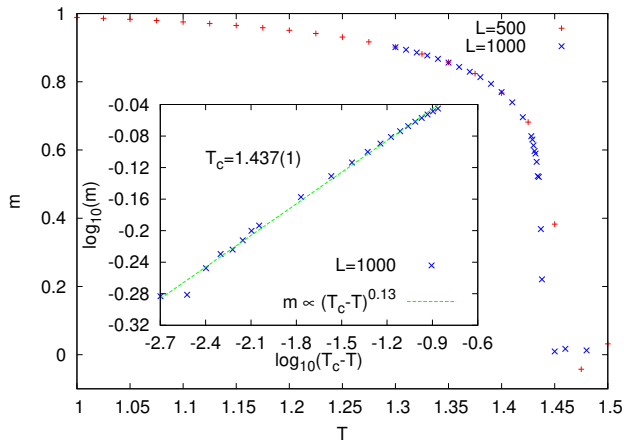}
\caption{(Color online) Magnetization $m$ as a function of temperature $T$ for the directed triangular lattice model. The estimated critical exponent $\beta=0.13(2)$ (inset) seems to be close to the equilibrium Ising model value $\beta=0.125$~\cite{yang}. Simulations started from the fully ferromagnetic configuration ($s_i=1$), and after $t=10^6$ skipped for relaxation, lasted for the same time interval.} 
\label{magnettriang}
\end{figure}

\begin{figure}
\includegraphics[width=\columnwidth]{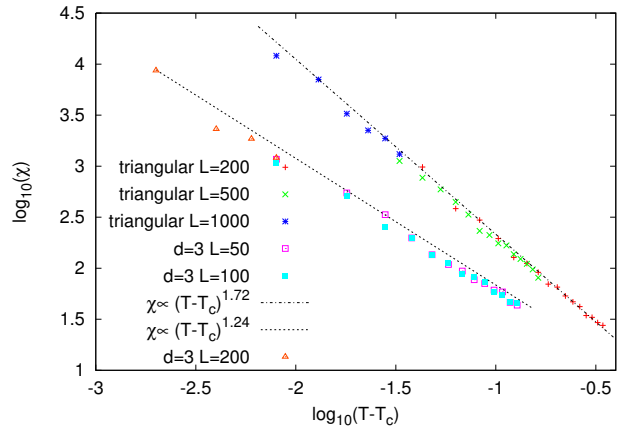}
\caption{(Color online) Susceptibility $\chi$ as a function of temperature $T(>T_c)$ for the Ising model on directed triangular and simple cubic ($d=3$) lattices. Estimated exponents $\gamma=1.72(3)$ (triangular) and $\gamma=1.24(2)$ (simple cubic) seem to be consistent with the equilibrium Ising model values $\gamma(d=2)=1.75$~\cite{baruch} and $\gamma(d=3)=1.2373(2)$~\cite{pelisetto}.}
\label{susc}
\end{figure}

It was suggested that the absence of the ferromagnetic ordering for the directed square lattice could be attributed to the violation of the action-reaction principle~\cite{lima2006}. Namely, on our directed lattices $s_i$ affects $s_j$ but not {\it vice versa}. Simulations for the directed triangular lattice (Fig.~\ref{lattice}b) question, however, such an explanation. Indeed, calculations of the magnetization (Fig.~\ref{magnettriang}) and susceptibility (Fig.~\ref{susc}) strongly suggest that the model undergoes a typical Ising-like phase transition with power-law singularities $m\sim (T_c-T)^{\beta}$ and $\chi \sim |T_c-T|^{-\gamma}$. What is more, the estimated exponents $\beta=0.13(2)$ and $\gamma=1.72(3)$ remain in a good agreement with the values for the two-dimensional Ising model, namely  $\beta=0.125$~\cite{yang} and $\gamma=1.75$~\cite{baruch}. We estimated the location of the critical point ($T=1.437(1)$) and values of critical exponents using the non-linear least-square method known as Marquardt-Levenberg algorithm \cite{nr}, that is implemented in some popular data-processing software like e.g., Gnuplot.

\subsection{d=3 simple cubic lattice}
\begin{figure}
\includegraphics[width=\columnwidth]{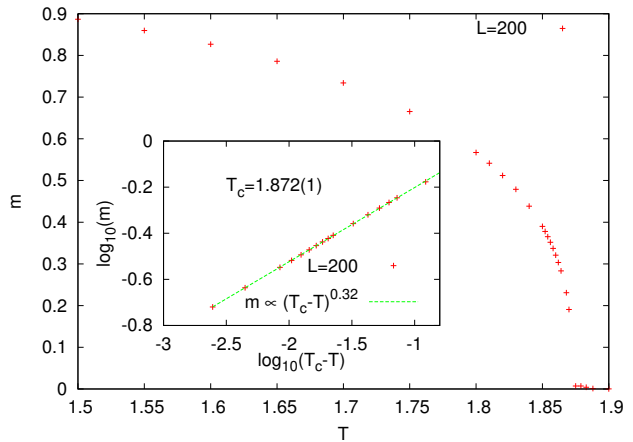}
\caption{(Color online) Magnetization $m$ as a function of temperature $T$ for the Ising model on the directed simple cubic lattice. The estimated critical exponent $\beta=0.32(2)$ (inset) is in a good agreement with the equilibrium Ising model value $\beta=0.3265(1)$~\cite{pelisetto}. Simulations start from the fully ferromagnetic configuration ($s_i=1$), and after $t=10^6$ skipped for relaxation, lasted for the same time interval.}
\label{magnetd3}
\end{figure}

We also performed simulations for the directed three-dimensional simple cubic lattice (Fig.~\ref{lattice}c; interacting neighbours: Up, Right and Back). The estimated values of critical exponents $\beta=0.32(2)$ (Fig.~\ref{magnetd3}) and $\gamma=1.24(2)$ (Fig.~\ref{susc}) show that also in this case the model most likely belongs to the equilibrium Ising model universality class, for which $\beta=0.3265(1)$ and $\gamma=1.2373(2)$~\cite{pelisetto}.

Our extensive simulations clearly indicate that for triangular and simple cubic lattices the model has a finite-temperature phase transition. Our results can be confronted with the Godr\`eche and Bray analyses of the Ising model (with the heat-bath dynamics) on directed Cayley tree~\cite{godreche}. They have shown that a finite-temperature transition occurs only for the branching number $q\geq 3$. Taking the number of out-links as the analogue of the branching number, we obtain that both triangle and simple cubic lattices have $q=3$ and thus Cayley-tree analysys support our numerical results. Let us also notice that for $q=3$  Godr\`eche and Bray obtain $T_c=2.1038...$, which is not much different from our simple-cubic result $T_c=1.872(1)$. 

\section{Directed random graphs}
We also analysed the behaviour of the Ising model with the heat-bath dynamics on directed random graphs. Such graphs are constructed using the following rule: for each directed pair $(i,j)$, with probability $p/N$ a link from $i$ to $j$ is created (and with probability $1-p/N$ the sites remain unconnected). The Ising model with heat-bath dynamics was already  examined on such graphs~\cite{limasumour}. It was shown that for $p=1$ the model is disordered at any positive temperature but for $p=2$ and 3 simulations suggest a positive $T_c$. Let us notice that $p=1$ corresponds to the percolation threshold on such graphs~\cite{barabasihavlin,dorog2001}. Before we briefly sketch the derivation of this well-known result, let us recall that percolation on directed networks is more subtle than on undirected ones. In particular, the notion of a cluster to which a given site belongs must be modified so as to take into account directedness of links. Thus, one introduces the notion of an in-component defined as a set of sites from which a given site can be reached, and of an out-component being a set of sites reachable from this site. Of interest is also a strongly connected component, where from each site one can reach any other site of this component. To examine the percolative  properties of these components, one might use the generating function $\Phi(x,y)$~\cite{dorog2001} defined as 
\begin{equation}
\Phi(x,y)=\sum_{k_i,k_o} P(k_i,k_o)x^{k_i}y^{k_o},
\label{gen-funct}
\end{equation}
 where $P(k_i,k_o)$ is the probability that a given  site has $k_i$ in-links and $k_o$ out-links. For our random graphs in the limit $N\rightarrow\infty$ we have
\begin{equation}
P(k_i,k_o)=\frac{p^{k_i}e^{-p}}{k_i!}\frac{p^{k_o}e^{-p}}{k_o!}.
\label{prob}
\end{equation}
Using Eq.~(\ref{prob}), one can easily calculate the generating function (\ref{gen-funct}) and obtain
\begin{equation}
\Phi(x,y)=e^{p(x-1)}e^{p(y-1)}.
\label{phi}
\end{equation}
Knowing the generating function $\Phi(x,y)$, one then calculates the generating function of the in-link distribution of the site which can be reached moving  against the link direction 
\begin{equation}
\Phi_1^{(i)}(x)=\frac{1}{p}\frac{\partial}{\partial y}\Phi(x,y)\bigg|_{y=1} 
\end{equation}
whose derivative being equal to unity sets the condition for the divergence of the in-component 
\begin{equation}
\frac{\partial}{\partial x}\Phi_1^{(i)}(x)\bigg|_{x=1}=1.
\label{divergence}
\end{equation}
Using Eq.~(\ref{phi}), one obtains $\Phi_1^{(i)}(x)=e^{p(x-1)}$ and from the condition (\ref{divergence}) we obtain that $p=1$ determines the point of divergence of the size of the in-component. Analogously, one finds the generating function of the out-link distribution of the site which can be reached moving  along the link direction $\Phi_1^{(o)}(y)=\frac{1}{p}\frac{\partial}{\partial x}\Phi(x,y)\big|_{x=1}=e^{p(y-1)}$ and from the condition $\frac{\partial}{\partial y}\Phi_1^{(o)}(y)\big|_{y=1}=1$ one finds that again $p=1$ marks the divergence of the size of the out-component. For factorizable probability distributions $P(k_i,k_o)$ (as, e.g., in Eq.~(\ref{prob})), the divergence of the strongly connected component is related to the divergence of in- and out-components~\cite{dorog2001}. In our case it means that $p=1$ marks also the divergence of the strongly connected component.

\begin{figure}
\includegraphics[width=\columnwidth]{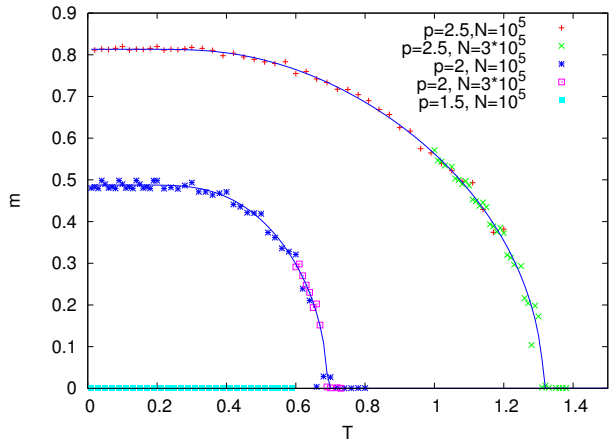}
\caption{(Color online) Magnetization $m$ as a function of temperature $T$ for the Ising model on the directed random graph. For $p=1.5$ (which is above percolation threshold $p=1$) even at $T=0.01$ the ferromagnetic initial configuration quickly becomes disorderd ($m=0$). The simulation and relaxation times were $t=10^5$. The continuous blue line shows a numerical solution of approximation (\ref{mfa}).}
\label{magnet}
\end{figure}

One can thus ask whether for $p>1$ the percolating components can support finite-temperature ferromagnetic ordering. Such situation takes place for the Ising model on undirected random graphs~\cite{replica,dorog}. Our simulations show that for directed graphs this is not the case (Fig.~\ref{magnet}). Indeed, while for $p=2.5$ and 2 we can observe a finite-temperature ordering, for $p=1.5$, which is above the percolation threshold ($p=1$), the model is not ordered at any positive temperature. Let us notice that for  $p=2$ and 2.5 and at low temperatures, the magnetization remains much smaller than unity, even though the simulations start from the ferromagnetic configuration (all spins $s_i=1$).

\begin{figure}
\includegraphics[width=\columnwidth]{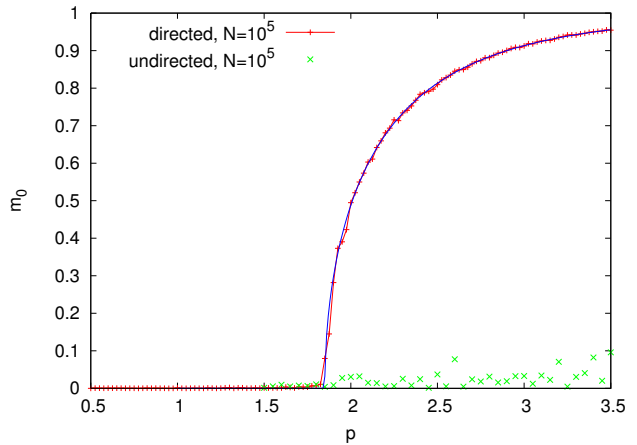}
\caption{(Color online) Absolute value of a zero-temperature magnetization $m$ as a function of $p$ for the Ising model on the directed and undirected random graph. The system starts at a random configuration of spins and evolves until a stationary state is reached. While for the undirected random graph, the model always ends up in the disordered state $m\sim 0$, for the directed graph and for $p>1.9$, a broken symmetry magnetized attractor is reached. The continuous blue line shows a numerical solution of approximation (\ref{mfa}).}
\label{magnet_t0}
\vspace{-0.3cm}
\end{figure}

Such behaviour inclined us to examine whether a ferromagnetic state (with a spontaneously broken symmetry) is actually the attractor of the zero-temperature dynamics on such graphs. Thus, we ran  $T=0$ simulations that started from the random initial configuration and we monitored the magnetization of the  final configuration. Typically, the model quickly approaches the final state and $p$-dependence of the zero-temperature magnetization is presented in Fig.~\ref{magnet_t0}. One can notice that, indeed, the ferromagnetic state is an attractor of the dynamics but only for $p>1.9$. For smaller values of $p$ (but still above the percolation threshold), the dynamics keeps the system in a paramagnetic ($m=0$) state. (We monitored also the total number of ferromagnetically oriented bonds, so that even in the paramagnetic case we could decide that the system reached the stationary state.) It is thus not surprising that we were not able to observe a finite-temperature ferromagnetism for $p<1.9$, since in such a case even at zero temperature the system is paramagnetic. Let us notice that such a zero-temperature coarsening is much different in undirected random graphs, where the attractor is paramagnetic for any~$p$~\cite{castellano}. Our simulations in the case of undirected random graphs confirm such a behaviour (Fig.~\ref{magnet_t0}).

Only close to $p=1.9$, we found that the approach to the steady state is much slower and perhaps power law. We expect that at this value the model might exhibit some kind of a critical behaviour, but we will not present its more detailed analysis here. 

In the final part of this section we would like to show that the behaviour of the Ising model on the directed random graph can be reproduced using a certain approximate scheme.  Our intention is to write an evolution equation for the probability $P(t)$ that a given spin, say $s_i$, takes the value +1 at time~$t$. According to the dynamics of the model (\ref{heat-bath}),  the probability $P(t)$ depends on the spin configurations of out-neigbours of~$i$. Let us notice, however, that since the graph is rare, those  out-neighbours are most likely  connected/correlated neither with each other nor with~$i$. Provided that the out-neighbours of the site~$i$ are described by the same probability as $s_i$, we obtain that $P(t)$ obeys the following equation
\begin{equation}
\frac{\partial P(t)}{\partial t}=-P(t)+\sum_{l=0}^{\infty} \frac{p^le^{-p}}{l!} R_l(t),
\label{mfa}
\end{equation}
where 
\begin{equation}
R_l(t)=\sum_{k=0}^{l} {l \choose k} \frac{P^k(t)[1-P(t)]^{l-k}}{1+\exp{[-4(k-l/2)/T]}}.
\label{mf1}
\end{equation}
In Eq.~(\ref{mfa}) the term $\frac{p^le^{-p}}{l!}$ is the probability that a randomly chosen site has $l$ out-links and $R_l(t)$ is the probability that such a site is to be set to +1. The steady state of Eq.~(\ref{mfa}) can be easily found numerically, by equating the r.h.s. to~0 and cutting-off the infinite summation over $l$ at a certain $l_{max}$ (we used $l_{max}=20$ but larger values give nearly the same results). The numerical results, shown in Figs.~\ref{magnet}--\ref{magnet_t0}, are in an excellent agreement with the Monte Carlo data.

Although very accurate, the approximation (\ref{mfa}) is not exact. For some values of $p$ and $T$, we performed more extensive simulations, which show that this approximation in some cases has an error $\sim 1\%$. We attribute such a disagreement mainly to the neglect of heterogeneity in (\ref{mfa}) but more detailed analysis of the accuracy of this kind of approximations is rather difficult \cite{mendesreview}. More accurate description of our model should use a set of degree-dependent functions $P_l(t)$ instead of a single function $P(t)$, but the resulting scheme would be numerically much more demanding.

\section{Out-homogeneous random graphs}
In our opinion, it is interesting to discuss the behaviour of the Ising model on yet another class of directed networks, which in a way interpolates between regular networks (Section II) and random graphs (Section III). Such networks are constructed in the following way: for each site of a network, we generate $z$ out-links to sites randomly selected from the remaining ones (multiple links to the same site are forbidden). Of course, for such networks the distribution of in-links is Poissonian as for random graphs. Since each site has a constant number of out-links, we call such a graph out-homogeneous. The merit of the analysis on such networks is that both magnetic and percolative properties can be examined (we believe) exactly. The Ising model on such  networks was used as a simple model of a financial market, where each spin $s_i$ represents an agent, whose decision to buy ($s_i=1$) or sell ($s_i=-1$) is influenced by the decisions of its $z$~neighbours~\cite{gontarek}.

In the out-homogeneous case, the evolution equation for $P(t)$ is obtained replacing  Poissonian distribution in Eq.~(\ref{mfa}) with $\delta_{l,z}$
\begin{equation}
\frac{\partial P(t)}{\partial t}=-P(t)+ \sum_{k=0}^{z} {z \choose k} \frac{P^k(t)[1-P(t)]^{z-k}}{1+\exp{[-4(k-z/2)/T]}}.
\label{gont}
\end{equation}
The steady-state predictions of this equation were already analysed and accurate Monte Carlo simulations strongly suggest that Eq.~\ref{gont}) provides an exact description of the Ising model~\cite{gontarek}. This is not surprising since, as we already argued, due to a constant number of out-links,  there is no problem of heterogeneity and a description in terms of a single probability function $P(t)$ should be sufficient.

Actually, in this case the graph locally becomes equivalent to the directed Cayley tree, and the Ising model with the heat-bath dynamics was already analysed on such a network by Godr\`eche and Bray~\cite{godreche}. They found that there is a ferromagnetic ordering in the system but only for $z>2$ while for $z=1$ and 2 the critical temperature $T_c=0$. Our Eq.~(\ref{gont}) is equivalent to that obtained by Godr\`eche and Bray, and in particular, from that equation one can easily obtain the critical temperature  for $z=1, 2, 3$ and~4 \cite{gontarek} that is in agreement with that already found by Godr\`eche and Bray.

One can also examine percolative properties of the out-homogeneous networks. 
In this case the probability distribution $P(k_i,k_o)$ has the form
\begin{equation}
P(k_i,k_o)=\delta_{k_o,z}\frac{z^{k_i}e^{-z}}{k_i!},
\label{prob-homo}
\end{equation}
where $\delta$ is the Kronecker delta function.
For such distribution, we obtain that the generating function  (\ref{gen-funct}) becomes $\Phi(x,y)=x^ze^{z(y-1)}$. Repeating the calculations of the previous section, we find $\Phi_1^{(i)}(x)=e^{z(x-1)}$ and $\Phi_1^{(o)}(y)=y^z$. Then we easily obtain that $z=1$ marks the percolation transition of in-, out-, and strongly connected components.

Thus, for $z=2$  the model is deeply in the percolating phase with infinite in-, out-, and strongly connected components,  but this is not enough to support the ferromagnetic ordering since in this case $T_c=0$.

\section{Conclusions}
In the present paper we examined the Ising model on several directed networks. The behaviour on regular networks suggests that such models have much in common with the equilibrium Ising models. However, the absence of the (finite-temperature) ferromagnetism on a square lattice and its presence on a triangular lattice show that the dimension and coordination number play here a more subtle role than in equilibrium models and some more general understanding of this aspect would be desirable. 

In a way, our results on heterogeneous directed random graphs lead to a similar conclusion.  Namely, while a percolating cluster provides a sufficient support for the finite-temperature ferromagnetism in undirected (i.e., equilibrium) Ising models, this is not enough for directed networks. Such an ordering appears but only for considerably denser networks (with larger~$p$). It would be interesting to examine whether the appearance of ferromagnetism in this case might be related to a certain topological change of the network such as, for example, percolation of some other (more complex) geometrical structures. 

Let us notice, however, that since our model does not obey the detailed balance, some other dynamics (Metropolis, cluster algorithms) might exhibit a different, but as we expect qualitatively similar, behaviour. Of interest would be to examine also directed network versions of some other closely related models such as, for example, the voter model~\cite{liggett}.

Acknowledgements: The research for this work was supported by NCN grant 2013/09/B/ST6/02277 (A.L.) and Ministry of Science and  Higher Education grant N N202 488039 (K.G.). This work was also partially funded by FEDER funds through the COMPETE 2020 Programme and National Funds throught FCT - Portuguese Foundation for Science and Technology under the project UID/CTM/50025/2013. 

\end {document}